\documentclass[twocolumn,nolinenumbers]{aastex701}
\usepackage[utf8]{inputenc}
\usepackage{amsmath}

\definecolor{newred}{rgb}{0.1,0.6,0.75}
\definecolor{grey}{rgb}{0.2,0.27,0.57}
\definecolor{freeblue}{rgb}{0.2,0.25,0.45}
\definecolor{freeblue2}{rgb}{0.2,0.25,0.4}
\definecolor{blue}{rgb}{0.45,0.41,0.98}
\definecolor{myblue}{rgb}{0.1,0.5,0.5}
\definecolor{guillaume}{rgb}{0.,0.5,0.65}
\definecolor{louloublue}{rgb}{0.2,0.2,0.65}
\definecolor{loulougreen}{rgb}{0.1,0.35,0.75}
\definecolor{violet}{rgb}{0.5,0.,0.5}

\newcolumntype{i}{>{\scriptsize}r}

\shortauthors{Readhead et al.}

\graphicspath{{./}{figs/}}

\begin{document}

\shorttitle{Confirmation of a Harmonic in the Blazar PKS J1309+1154}

\title{Compelling Evidence for a Harmonic in the Light Curve of the Supermassive Black Hole Binary Candidate PKS J1309+1154}
\correspondingauthor{Anthony Readhead}
\email{acr@caltech.edu}

\author[0000-0001-9152-961X]{A.~C.~S.~Readhead}
\email{acr@caltech.edu}
\affiliation{Owens Valley Radio Observatory, California Institute of Technology,  Pasadena, CA 91125, USA}

\author[0000-0003-2483-2103]{M. F. Aller}
\email{mfa@umich.edu}   % Margo Aller
\affiliation{Department of Astronomy, University of Michigan, 323 West Hall, 1085 S. University Avenue, Ann Arbor, MI 48109, USA}

\author[0000-0002-9545-7286]{A. G. Sullivan}
\email{andrew.sullivan@stanford.edu}   % Andrew Sullivan
\affiliation{Kavli Institute for Particle Astrophysics and Cosmology, Department of Physics,
Stanford University, Stanford, CA 94305, USA}

\author[0000-0002-1854-5506]{R. D. Blandford}
\email{rdb3@stanford.edu}   % Roger Blandford
\affiliation{Kavli Institute for Particle Astrophysics and Cosmology, Department of Physics,
Stanford University, Stanford, CA 94305, USA}

\author[0000-0001-7016-1692]{P. Mr{\'o}z}
\email{Pmroz@astrouw.edu.pl}   % Przemek Mroz
\affiliation{Astronomical Observatory, University of Warsaw, Al. Ujazdowskie 4, 00-478 Warszawa, Poland}

\author[0000-0001-5957-1412]{P. V.~De la Parra}
\email{phvergara@udec.cl}   % Philipe Alexis Vergara De La Parra
\affiliation{CePIA, Astronomy Department, Universidad de Concepci\'on,  Casilla 160-C, Concepci\'on, Chile}

\author[0009-0000-9963-6874]{B. Molina}
\email{brian.leftraru@gmail.com}   % Brian Molina
\affiliation{CePIA, Astronomy Department, Universidad de Concepci\'on,  Casilla 160-C, Concepci\'on, Chile}

\author[0000-0002-0491-1210]{E.R. Most}
\email{emost@caltech.edu}
\affiliation{TAPIR, Mailcode 350-17, California Institute of Technology, Pasadena, CA 91125, USA}
\affiliation{Walter Burke Institute for Theoretical Physics, California Institute of Technology, Pasadena, CA 91125, USA}

\author[0000-0003-1315-3412]{M. L. Lister}
\email{mlister@purdue.edu}   % Matt Lister
\affiliation{Department of Physics and Astronomy, Purdue University, 525 Northwestern Avenue, West Lafayette, IN 47907, USA}

\author[0009-0004-2614-830X]{A. Synani} 
\email{akyvsyn@physics.uoc.gr}   % Anna Synani
\affiliation{Department of Physics and Institute of Theoretical and Computational Physics, University of Crete, GR-70013 Heraklion, Greece}
\affiliation{Institute of Astrophysics, Foundation for Research and Technology-Hellas, GR-70013 Heraklion, Greece}

\author[0000-0003-1945-1840]{H. Aller}
\email{haller@umich.edu}   % Hugh Aller
\affiliation{Department of Astronomy, University of Michigan, 323 West Hall, 1085 S. University Avenue, Ann Arbor, MI 48109, USA}

\author[0000-0003-0936-8488]{M. C.  Begelman}
\email{mitch@jila.colorado.edu}
\affiliation{JILA, University of Colorado and National Institute of Standards and Technology, 440 UCB, Boulder, CO 80309-0440, USA} 

\author[0000-0002-5770-2666]{Y. Ding}
\email{yding@caltech.edu}   % "Ding, Yuanze (Zach)"
\affiliation{Cahill Center for Astronomy and Astrophysics, California Institute of Technology, Pasadena, CA 91125, USA}

\author[0000-0002-3168-0139]{M. J. Graham}
\email{mjg@caltech.edu}   % "Graham, Matthew J."
\affiliation{Division of Physics, Mathematics, and Astronomy, California Institute of Technology, Pasadena, CA 91125, USA}

\author[0000-0002-4226-8959]{F. Harrison}
\email{fiona@srl.caltech.edu}   % "Harrison, Fiona A."
\affiliation{Cahill Center for Astronomy and Astrophysics, California Institute of Technology, Pasadena, CA 91125, USA}

\author[0000-0002-2024-8199]{T. Hovatta}
\email{talvikki.hovatta@utu.fi}   % Talvikki Hovatta
\affiliation{Finnish Centre for Astronomy with ESO (FINCA), University of Turku, FI-20014 University of Turku, Finland}
\affiliation{Aalto University Department of Electronics and Nanoengineering, PL~15500, FI-00076 Espoo, Finland}
\affiliation{Aalto University Mets\"ahovi Radio Observatory,  Mets\"ahovintie 114, 02540 Kylm\"al\"a, Finland}

\author[0000-0001-9200-4006]{I. Liodakis}
\email{yannis.liodakis@gmail.com}   % Yannis Liodakis
\affiliation{Institute of Astrophysics, Foundation for Research and Technology-Hellas, GR-70013 Heraklion, Greece}

\author[0000-0002-5491-5244]{W. Max-Moerbeck} 
\email{wmax@das.uchile.cl}   % Walter Max-Moerbeck
\affiliation{Departamento de Astronomía, Universidad de Chile, Camino El Observatorio 1515, Las Condes, Santiago, Chile}

\author[0000-0002-0870-1368]{V. Pavlidou} 
\email{pavlidou@physics.uoc.gr}  % Vasiliki Pavlidou
\affiliation{Department of Physics and Institute of Theoretical and Computational Physics, University of Crete, GR-70013 Heraklion, Greece}
\affiliation{Institute of Astrophysics, Foundation for Research and Technology-Hellas, GR-70013 Heraklion, Greece}

\author[0000-0001-5213-6231]{T. J. Pearson}
\email{tjp@astro.caltech.edu}   % "Pearson, Timothy J."
\affiliation{Owens Valley Radio Observatory, California Institute of Technology,  Pasadena, CA 91125, USA}

\author[0000-0002-7252-5485]{V. Ravi}
\email{vikram@astro.caltech.edu}   %  "Ravi, Vikram"
\affiliation{Owens Valley Radio Observatory, California Institute of Technology,  Pasadena, CA 91125, USA}

\author[0000-0001-5704-271X]{R. A. Reeves}
\email{rreeves@udec.cl}  %  Rodrigo Reeves
\affiliation{CePIA, Astronomy Department, Universidad de Concepci\'on,  Casilla 160-C, Concepci\'on, Chile}

\author[0000-0002-6369-6266]{T. Surti}
\email{tsurti@caltech.edu}  % "Surti, Tirth D."
\affiliation{Owens Valley Radio Observatory, California Institute of Technology,  Pasadena, CA 91125, USA}

\author[0000-0002-8831-2038]{K. Tassis}
\email{tassis@physics.uoc.gr}   % Konstantinos Tassis
\affiliation{Department of Physics and Institute of Theoretical and Computational Physics, University of Crete, GR-70013 Heraklion, Greece}
\affiliation{Institute of Astrophysics, Foundation for Research and Technology-Hellas, GR-70013 Heraklion, Greece}

\author[0000-0001-7662-2576]{S. E. Tremblay}
\email{strembla@nrao.edu}   % Steven Tremblay
\affiliation{National Radio Astronomy Observatory, 1011 Lopez Road, Socorro, NM 87801, USA}

\author[0000-0001-7470-3321]{J. A. Zensus}
\email{azensus@mpifr-bonn.mpg.de}   % Anton Zensus
\affiliation{Max-Planck-Institut f\"ur Radioastronomie, Auf dem H\"ugel 69, D-53121 Bonn, Germany}

\begin{abstract}
We recently discovered  a supermassive black hole binary (SMBHB) candidate, PKS J1309+1154,  in the combined 46-yr University of Michigan Radio Astronomy Observatory (UMRAO) plus Owens Valley Radio Observatory (OVRO) blazar monitoring programs at 14.5/15 GHz. The light curve of PKS 1309+1154 exhibits a 17.9 year periodicity.  We also reported a hint of a first harmonic with a 9 year periodicity  in this object.  Further analysis of the PKS J1309+1154 light curve provides compelling evidence that both the fundamental and the harmonic are real, confirming the existence of real periodicities in blazar light curves. This is the first case, to our knowledge, of watertight evidence for a fundamental and a harmonic periodicity in a blazar light curve.  It makes PKS J1309+1154  a \textit{strong\/} supermassive black hole binary (SMBHB) candidate, and thus the third such candidate to be revealed through long-term radio monitoring, the other two being PKS J0805--0111 and PKS 2131--021, both discovered through the OVRO 40 m Telescope monitoring program.   It is argued that hundreds of SMBHB candidates will be discovered by the Vera Rubin and Simons Observatories. Coherent searches for gravitational waves from a network of SMBHB candidates, starting immediately, are strongly motivated.
\end{abstract}


\section{Introduction}
\label{sec:intro}

The recently discovered evidence for a stochastic background of gravitational waves with periods of months to years \citep{2023ApJ...951L...8A,2023ApJ...952L..37A,2023ApJ...951L..50A,2023A&A...678A..50E,2023PASA...40...49Z,2025MNRAS.536.1489M,2025arXiv250816534A} relies on millisecond pulsar arrays for the timing,\footnote{There were two crucial steps in the discovery of millisecond pulsars: (i) the discovery of interplanetary scintillation, at Galactic latitude $-0.3^\circ$, in 4C 21.53 by \citet{1974MmRAS..78....1R},  which first drew attention to the singular nature of this object,  see \citet{2024JAHH...27..453R}; and (ii) the discovery of millisecond pulses from 4C 21.53W by \citet{1982Natur.300..615B}.} and is thought to be due to supermassive black hole binaries (SMBHBs) \citep{2023ApJ...952L..37A,2025arXiv250816534A}. This has spurred searches for periodicities in blazar light curves, which have the potential of revealing important clues to the dynamics of the central engines, powered by a supermassive black hole (SMBH), in active galactic nuclei (AGN) \citep{1984RvMP...56..255B}. The phenomenology of AGN, as revealed through their light curves, has been a challenging area of study because the expected timescale of any periodicities arising from the dynamics of their central engines ranges from months to decades.  This is due to the fact that the masses of the SMBHs in blazars are typically $10^7M_\odot-10^{10}M_\odot$ \citep{2019ARAandA..57..467B}, and very close binaries are unlikely to be seen because, due to gravitational radiation,  they spend  little time at small separations \citep{1980Natur.287..307B}.   

Until recently, the most compelling case has been that of  OJ 287 \citep[e.g.,][]{1996ApJ...460..207L,2007ApJ...659.1074V,2008Natur.452..851V,2013MNRAS.434.3122P,2016ApJ...819L..37V,2018MNRAS.478.3199B,Valtonen2023}. OJ 287 is, in many respects, a typical blazar. At radio, optical, and gamma-ray energies it exhibits bright flares, of duration ranging from weeks to several months, that dominate its light curve \citep{1998ApJ...503..662H,2004AandA...419..485C,1996ApJ...460..207L,2017A&A...597A..80H}. 

Our understanding of the phenomenology of blazar light curves is now changing owing to the large ($\sim 1830$ blazars), high-cadence (3--7 days), long-term (2008--present),  15 GHz monitoring campaign of the 40 m Telescope of the Owens Valley Radio Observatory (OVRO) \citep{2011ApJS..194...29R} and the mid 1970s--2012.5 monitoring campaign of the University of Michigan Radio Astronomy Observatory (UMRAO) \citep{1985ApJS...59..513A,2014ApJ...791...53A}, which has 83 frequently-observed blazars in common with the OVRO sample.  The blazars in common in these two samples  have therefore been monitored at high cadence for 46 years. These two data sets are unparalleled in numbers, cadence, and duration, for studying long term variations in blazars.

These monitoring campaigns have revealed that not all blazar light curves are dominated by short-lived flares of duration of weeks to months.  Some are dominated by strong slowly-varying  features, of months to years duration, upon which shorter-lived smaller amplitude flux density variations are imprinted. 

In the case of PKS 2131--021, \citet[hereafter Paper 1]{2022ApJ...926L..35O} and \citet[hereafter Paper 2]{2025ApJ...985...59K} have shown that coherent 4.75-yr sinusoidal variations, which cannot be ascribed to the steep spectrum (the ``red noise tail'') of the power spectral density (PSD) of the variations, are observed from 2.7 GHz to optical frequencies.

In PKS J0805--0111, \citet{2025ApJ...987..191D} and \citet{2025arXiv250404278H} have shown that strong coherent sinusoidal variations are present from 15 GHz to 225 GHz.  In addition, the combined UMRAO+OVRO monitoring campaign has recently discovered a third blazar, PKS J1309+1154 \citep[hereafter Paper 3]{2025arXiv251023103M}, with a light curve dominated  by coherent sinusoidal variations of period 17.9 yr from 15 GHz to 104 GHz. Furthermore, in addition to the strong sinusoidal emission,  evidence of a possible harmonic was found in the radio light curve of PKS J1309+1154.

The dominant variations in these three blazars are remarkably sinusoidal, and have been explained by the ``kinetic orbital'' or ``KO'' model in which the periodicity is ascribed to orbital motion in a supermassive black hole binary (SMBHB) (\citealt{2017MNRAS.465..161S} and Paper 1).  Further, more self-consistent modeling of the interior jet emission has successfully explained the multiwavelength phenomenology of PKS 2131--021 and PKS J0805--0111 \citep[hereafter Paper 4]{2025arXiv251002301S}. 

\begin{figure*}[t]
   \centering
   \includegraphics[width=0.9\linewidth]{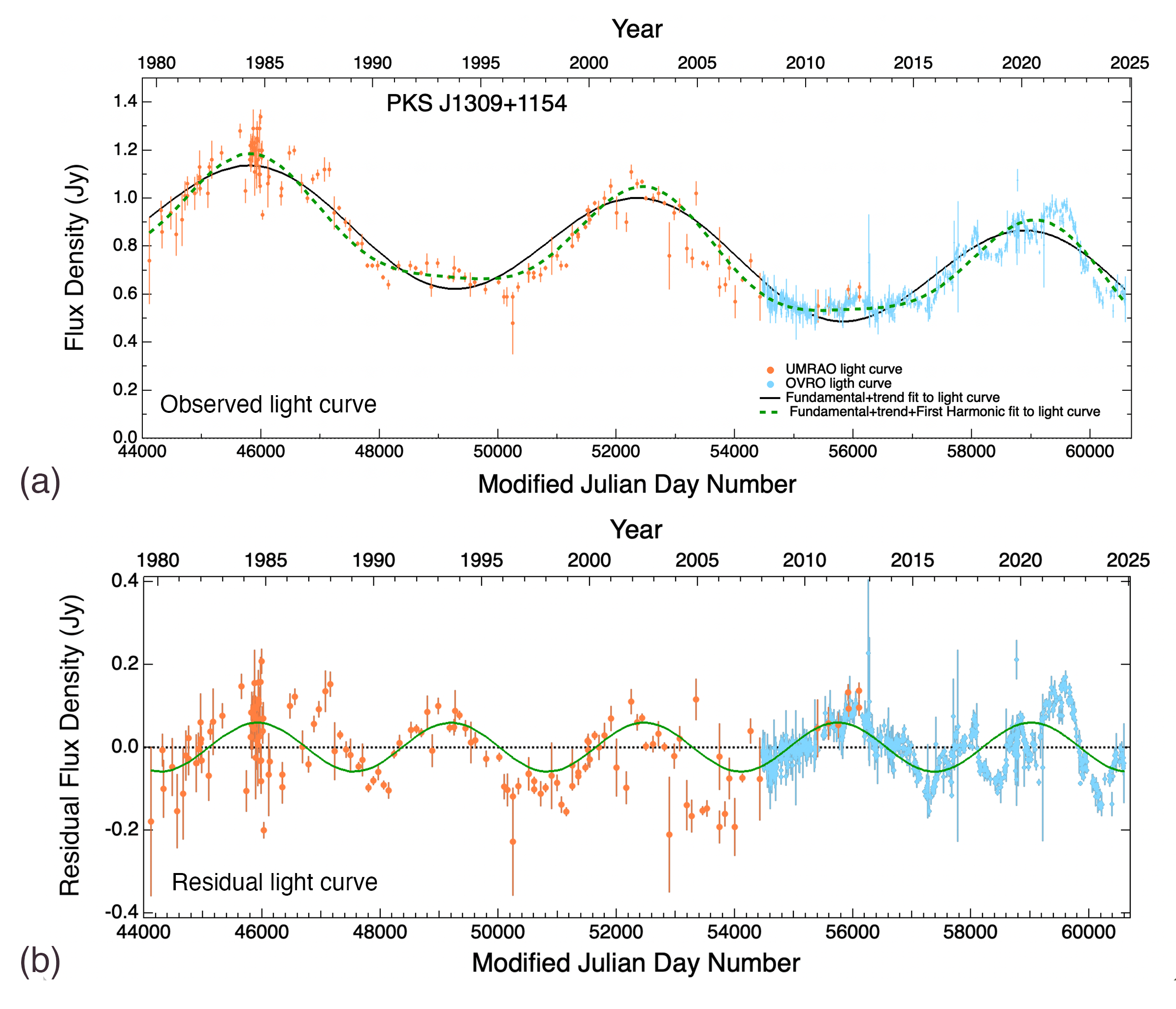}
   \caption{The PKS J1309+1154 fundamental and harmonic periodicities. Orange and blue symbols denote the UMRAO and OVRO data, respectively. (a) shows the observed light curve, (b) shows the residual light curve after subtraction of the black curve in (a), which is the least squares sine wave plus trend fit to the observed light curve.  The solid green curve in (b) shows the fitted least squares sine wave to the residual light curve, which has a period of the first harmonic of the fundamental (see text).  The green dashed  line in (a) shows the combined fundamental+trend+harmonic. }
   \label{plt:J1309harmonic}
\end{figure*}

\begin{deluxetable*}{c@{\hskip 8mm}cccccc}[!ht]
\tablecaption{Cycles, Sections, and Regions, of the Observations}
\tablehead{Cycle	&	Section	& Region&	MJD	&	MJD	&	Date	&	Date	\\
	&		&	&Start	&	End 	&	Start	&	End	
}
\startdata
0	&	4	&1&	43619	&	45256	&	4/21/78	&10/14/82\\
1	&	1	&2&	45256	&	46894	&	10/14/82	&4/9/87\\
1	&	2	&3&	46894	&	48531	&	4/9/97	&10/2/91\\
1	&	3	&4&	48531	&	50169	&	10/2/91	&3/27/96\\
1	&	4	&5&	50169	&	51806	&	3/27/96	&9/19/00\\
2	&	1	&6&	51806	&	53444	&	9/19/00	&3/15/05\\
2	&	2	&7&	53444	&	55081	&	3/15/05	&9/7/09\\
2	&	3	&8&	55081	&	56719	&	9/7/09	&3/3/14\\
2	&	4	&9&	56719	&	58356	&	3/3/14	&8/26/18\\
3	&	1	&10&	58356	&	59994	&	8/26/18	&2/19/23\\
3	&	2	&11&	59994	&	61631	&	2/19/23	&8/14/27\\
\enddata
\tablecomments{The cycles are of duration equal to the period of the fundamental (6550 days). The sections are of duration one half the period of the harmonic --- i.e., one quarter the period of the fundamental (1637.5 days). Thus there are eleven separate regions of the light curve referred to in this paper.}
\label{tab:cycles}
\end{deluxetable*}

In this paper we carry out an intensive study of the 46-year 14.5/15 GHz light curve of PKS J1309+1154. We provide compelling evidence to reject the null hypothesis that the light curve of PKS J1309+1554 does not  contain the harmonic found in Paper 3.  We believe that this is the first time that compelling evidence for a harmonic has been found in the light curve of a blazar. This is vitally  important because the existence of a harmonic constitutes proof of the reality of the fundamental. It is important to note that the evidence for this harmonic is \textit{not} based on the generalized Lomb-Scargle (GLS) spectrum analysis \citep{1976Ap&SS..39..447L,1982ApJ...263..835S,2009AandA...496..577Z}, which was shown using simulations in  in Paper 3, to produce an average of 4 non-existent harmonics per simulation. We therefore conclude that, in view of the harmonic we have detected,  PKS J1309+1154  is a \textit{strong\/} SMBHB candidate. The discovery of a harmonic in a blazar light curve greatly increases the predictive power of searches for  SMBHB candidates in  blazar light curves.

For consistency with our other papers, we assume the following cosmological parameters: $H_0 = 71$\, km\,s$^{-1}$\, Mpc$^{-1}$, $\Omega_{\rm m} = 0.27$, $\Omega_\Lambda = 0.73$  \citep{2009ApJS..180..330K}.  None of the conclusions would be changed were we to adopt the best model of the Planck Collaboration  \citep{2020AandA...641A...6P}.

\begin{figure*}
   \centering
   \includegraphics[width=1\linewidth]{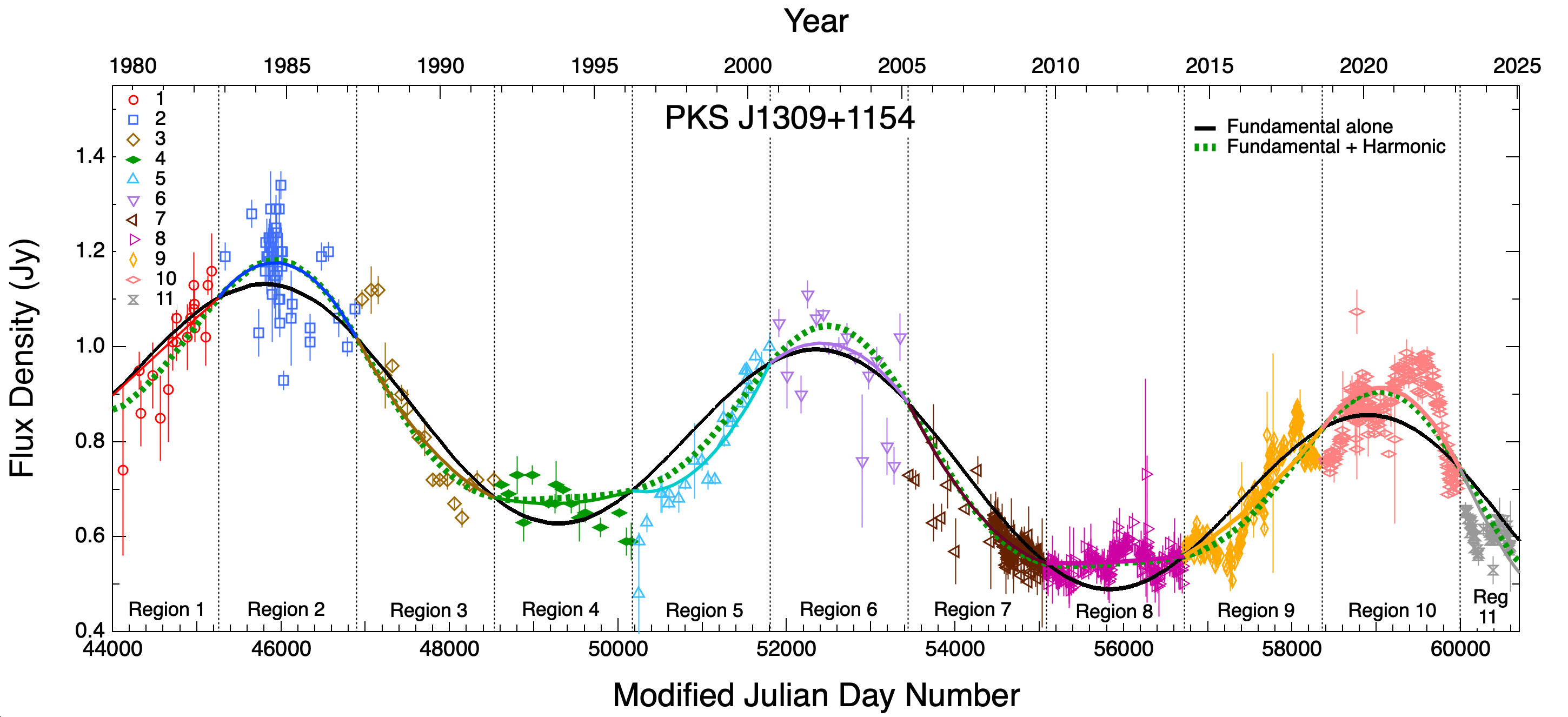}
   \caption{The PKS J1309+1154 light curve analyzed in the eleven distinct regions where the harmonic crosses zero, demarcated by the vertical dashed lines.  In each region a quadratic polynomial has been fitted to the light curve in that region anchored at the crossing points. In many cases the quadratic fit lies so close to it that it is indistinguishable from the fundamental+harmonic fit. In all eleven regions the quadratic fit lies on the same side of the fundamental fit alone as the fundamental+harmonic fit. We therefore reject the hypothesis that the data is randomly distributed relative to the fundamental, and, consequently, that the harmonic is not real, at the level  p-value $=4.9\times 10^{-4}$ --- i.e., 3.3$\sigma$.  }
   \label{plt:J1309regions}
\end{figure*}

\section{The Observations}\label{sec:observations}

The University of Michigan Radio Astronomical Observatory (UMRAO) carried out a blazar monitoring program on the UMRAO 26 m Telescope from the mid 1970s to 2012.5 \citep{1985ApJS...59..513A,2014ApJ...791...53A}.  Dual linearly-polarized feed horns symmetrically straddling  the prime focus fed a broadband receiver with bandwidth 1.68 GHz. An on--on observing technique alternated the beams on the source in order to compensate for the ground spillover and atmospheric effects.

The OVRO 40 m Telescope monitoring program was started in 2008 and has run continuously since that time.  The telescope has a dual, symmetric beam system. Before 2014 the two beams were Dicke-switched.  In 2014 the Dicke-switched receiver was replaced by a correlation receiver.  The dual-switching system, which greatly reduces atmospheric and ground spillover effects, has been described by \citet{1989ApJ...346..566R}, and details of the 40 m monitoring program are given in \citet{2011ApJS..194...29R}.

\begin{figure*}[t]
   \centering
   \includegraphics[width=0.7\linewidth]{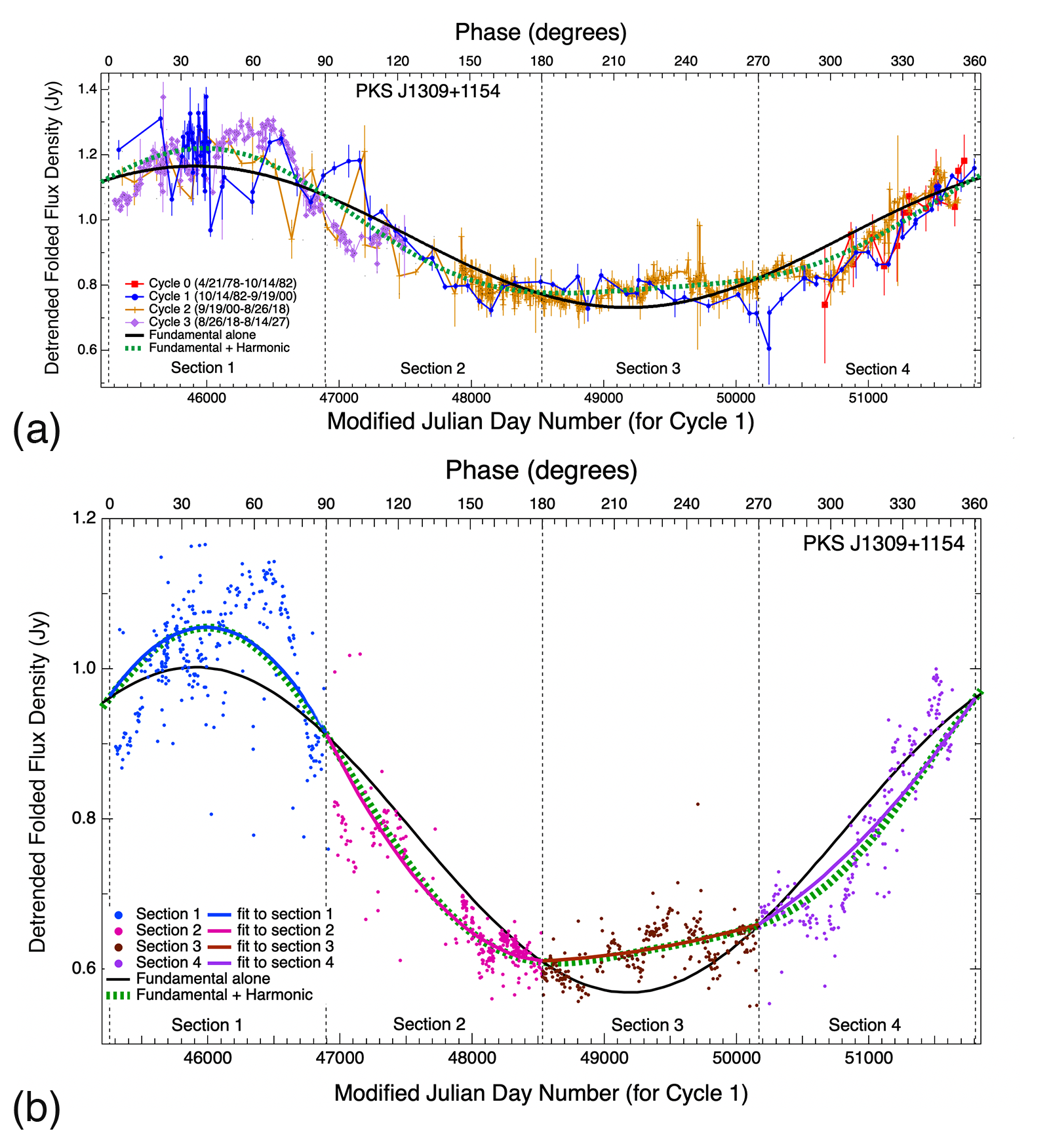}
   \caption{The detrended, folded light curve of PKS J1309+1154, assuming a period of 6550 days for the fundamental periodicity. (a) The folded light curve showing the different cycles (see Table \ref{tab:cycles}). Here we have folded cycles 0, 2, and 3 on to cycle 1.  (b) Here the folded data have been divided into the four sections listed in Table \ref{tab:cycles}. Least squares fits of quadratic polynomials have been performed on each section separately, with the polynomials anchored at the zero points of the harmonic, shown here by the crossover points of the fundamental \textit{vs.\/} the fundamental+harmonic fits. Rather surprisingly these quadratic fits now lie almost identically on the fit of the fundamental + harmonic, showing clearly that the data are far more consistent with the fit including the harmonic rather than that excluding the harmonic. Comparison of the quadratic polynomial fits in Fig.~\ref{plt:J1309regions} compared to those shown in (b) above, shows that some of these fits are significantly better in the folded light curve, indicating the coherence of the underlying fundamental+harmonic signal over the 2.5-cycle duration of the observations.}
   \label{plt:folded}
\end{figure*}

\begin{figure*}[t]
   \centering
   \includegraphics[width=1\linewidth]{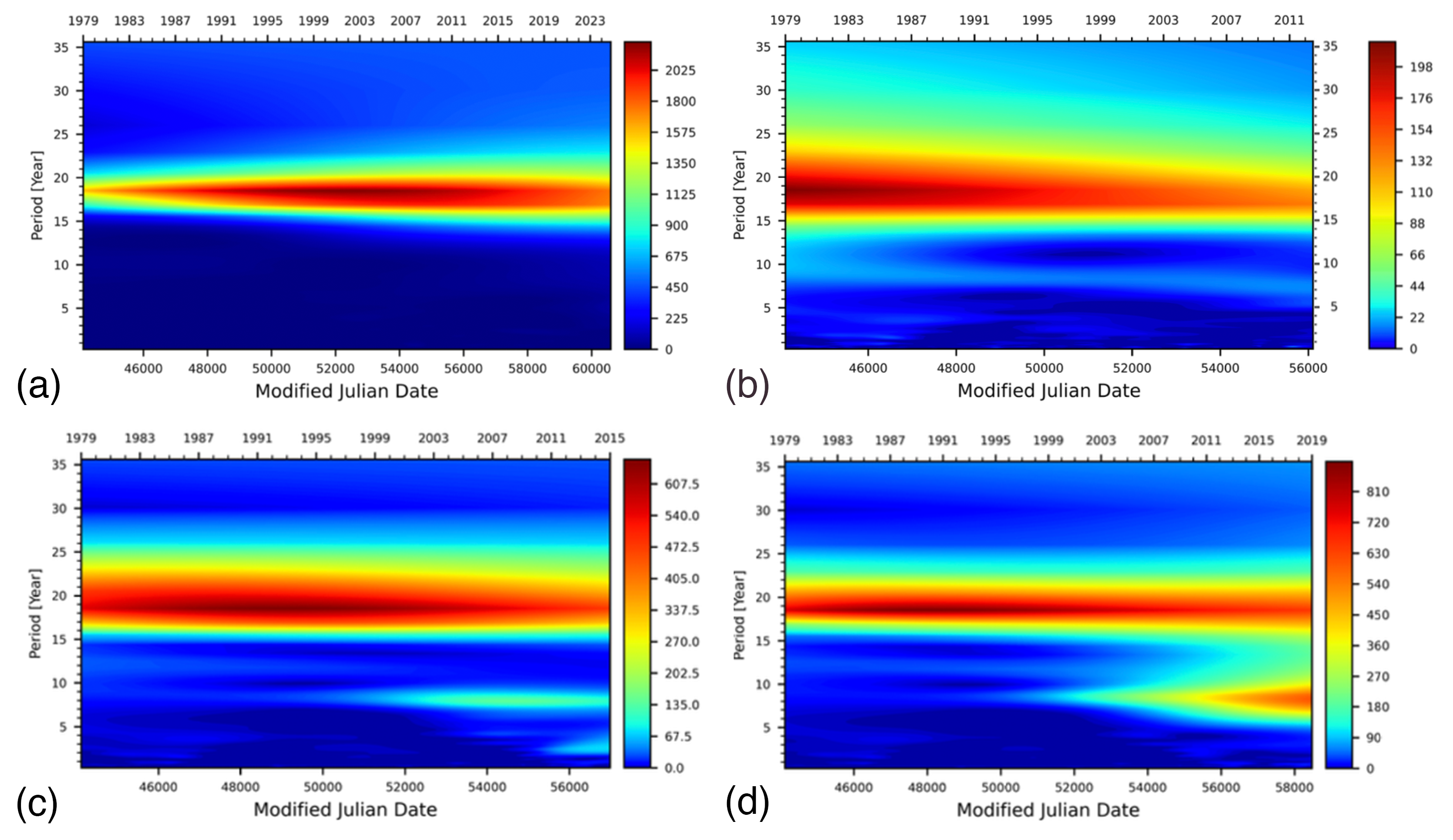}
   \caption{WWZ spectra of PKS J1309+1154 showing the fundamental periodicity and the first harmonic, which are both coherent over the whole light curve (see text).  (a) This is the spectrum of the full UMRAO+OVRO 46-yr light curve on the conventional linear power scale. (b)-(d) Show WWZ spectra using the square root of the power (see text). (b) Here we show the spectrum of the full (1979-2012.5) UMRAO light curve alone. (c) The spectrum of the full UMRAO light curve from 1979--2012.5 plus the OVRO light curve from 2012.5--2015. (d)  The spectrum of the full UMRAO light curve from 1979--2012.5 plus the OVRO light curve from 2012.5--2019.
 }
   \label{plt:J1309wwz}
\end{figure*}

\section{The Fundamental and First Harmonic Periodicities in PKS J1309+1154}\label{sec:J1309}

The combined UMRAO+OVRO 14.5/15 GHz light curve of PKS J1309+1154 is shown in Fig.~\ref{plt:J1309harmonic}(a). The UMRAO and OVRO  wide-band receivers provide considerable overlap in frequency.  The least squares fit to the light curve of a sine wave plus trend (black curve) has a (fundamental) period of 6551 days and a trend of $-0.00755$ Jy yr$^{-1}$. When a harmonic is included in the fit, the values are slightly different (period 6550 days, trend $-0.00769$ Jy yr$^{-1}$).   The fitting was done using the maximum likelihood method described in detail in Papers 1 and 2, and summarized in Appendix \ref{sec:sine}. An interesting feature of the fit without the harmonic is that the data clearly oscillate about the fundamental plus trend fit shown by the black curve  twice per cycle.  The residual light curve after subtracting the fundamental plus trend is shown in Fig.~\ref{plt:J1309harmonic}(b). The least squares sine wave fit to the residual light curve is shown by the green solid curve in Fig.~\ref{plt:J1309harmonic}(b).  In  Fig.~\ref{plt:J1309harmonic}(a) the heavy dashed green curve shows the sum of the fundamental sine wave plus trend plus the harmonic. In Paper 3 it was suggested only that there was a hint of a harmonic, and it was pointed out that the frequency of the harmonic is twice that of the fundamental, and also that the harmonic is in phase with the fundamental.  

In  Paper 3 we carried out our standard (see Paper 1) significance test based on simulations and showed that the fundamental is significant at the $2.3\sigma$ level.  This is the \textit{global} significance, i.e. the significance given that we had no \textit{a priori\/} knowledge of the period (see Paper 1). We also hypothesized in Paper 3, that the light curve of PKS J1309+1154 has a harmonic of twice the frequency in phase with the fundamental.  We have now carried out the same statistical analysis on the residual of the PKS J1309+1154 light curve after subtracting the fundamental plus trend, shown in  Fig. \ref{plt:J1309harmonic}(b). The local significance of the harmonic,  based on the fact that we are looking for a signal of known frequency, is $2.3\sigma$, and the alternate peaks of the harmonic align to within 10\% of a half-cycle of the harmonic.  These numbers strongly suggest that the harmonic is real, but they are not compelling.

We now examine the light curve of PKS J1309+1154 in detail. In Table \ref{tab:cycles} we show the terms we use for the different data sets we consider and their corresponding MJD ranges and dates.  we assume a fundamental period of 6550 days.  Over the 46-yt duration of the light curve the fundamental went through 2.57 ``cycles''. In principle we could start our first cycle at the beginning of the observations in  1979, but it simplifies the analysis to start the first cycle at the first zero of the harmonic at MJD 45256 (10/14/82).  We therefor label the partial cycle at the beginning ``cycle 0'',  In each cycle, there are  four ``sections'', equal to half the period of the harmonic, of length 1637.5 days.  This given us eleven separate ``regions'' where the harmonic lies alternately on one side of the fundamental or the other.

In Fig.~\ref{plt:J1309regions} we show the light curve of PKS J1309+1154 with the eleven regions demarcated by the zero crossings of the harmonic indicated by the vertical black dashed lines. Three very interesting features of this fit of the fundamental + harmonic are:

\begin{enumerate}

\item The harmonic creates a flat portion, lasting about 8 years, i.e., 45\% of the fundamental cycle, straddling the minima in the fundamental. This flat portion was observed from 1990--1998 and again from 2008--2016.  It is thus predicted to appear again from 2026--2034.

\item The harmonic, in addition to flattening out the minima in the light curve, creates higher peaks at the maxima.

\item In all 11 regions, throughout the whole light curve, the data alternates from side to side relative to the  fundamental plus trend fit (black curve) as time progresses.  We have fitted quadratic polynomials, anchored to the fundamental plus trend at the zero points of the harmonic, to the data in each of the 11 regions. The parameters of theses fits are given in Table \ref{tab:quadratic}.  The resulting fits are shown by the solid curves in the same color as the data in each region. As can be seen, these fits lie very close to the fundamental plus trend plus harmonic curve given by the green dashed line.  The random probability of the data lying on the same side of the fundamental plus trend as the harmonic in all eleven regions has p-value =$4.9 \times 10^{-4}$, so this result is significant at the $3.3\sigma$ level.

\end{enumerate}

Taken together, we find that these three points constitute compelling proof that the harmonic is real.  This, in turn, proves that the fundamental is real, and that these two periodicities in the light curve of PKS J1309+1154 are not a random result caused by the steep slope of the PSD. These three effects all result from the fact that the harmonic has frequency twice that of the fundamental and is in phase with the fundamental.

\subsection{The Folded light curve of PKS J1309+1154}\label{sec:folded}

  In Fig.~\ref{plt:folded}(a) we show the folded light curve, based on the period of 6550 days, in which the light curves from cycles 0, 2, and 3 have been folded on to that of cycle 1 (see Table \ref{tab:cycles}). In the folded curve, we delineate the  four sections of the fundamental period by the vertical dashed lines, which mark the time when the harmonic amplitude is zero (where the green curve crosses the black curve).  In Fig.~\ref{plt:folded}(a) the different cycles show clearly that the same pattern repeats in each cycle, with the main deviations from the light curve repeating from one cycle to the next. The oscillation of the light curve from one side of the fundamental sine wave fit to the other twice per period of the harmonic is very clear.  In Fig.~\ref{plt:folded}(b) we have fitted quadratic polynomials (see the solid color curves) anchored to the fundamental at the crossover points at the  section boundaries, which illustrates this oscillating behavior very clearly. The parameters of the quadratic fits are given in Table \ref{tab:quadratic2}. Note that the quadratic fits to the folded curve of Fig.
  In Fig.~\ref{plt:folded}(a) we show the folded light curve, based on the period of 6550 days, in which the light curves from cycles 0, 2, and 3 have been folded on to that of cycle 1 (see Table \ref{tab:cycles}). In the folded curve, we delineate the  four sections of the fundamental period by the vertical dashed lines, which mark the time when the harmonic amplitude is zero (where the green curve crosses the black curve).  In Fig.~\ref{plt:folded}(a) the different cycles show clearly that the same pattern repeats in each cycle, with the main deviations from the light curve repeating from one cycle to the next. The oscillation of the light curve from one side of the fundamental sine wave fit to the other twice per period of the harmonic is very clear.  In Fig.~\ref{plt:folded}(b) we have fitted quadratic polynomials (see the solid color curves) anchored to the fundamental at the crossover points at the  section boundaries, which illustrates this oscillating behavior very clearly. Note that the quadratic fits to the folded curve of Fig.~\ref{plt:folded}(b) are better in many cases than the quadratic fits of Fig.~\ref{plt:J1309regions}, indicating that the coherence of the harmonic continues from one cycle to the next throughout the light curve.

\subsection{The WWZ Spectrum and the Duration of the harmonic}\label{sec:duration}

We have also used the weighted wavelet Z-transform (WWZ) \citep{1996AJ....112.1709F} to analyze the light curve of PKS J1309+1154.
In Fig.~ \ref{plt:J1309wwz}(a)-(d) we show WWZ spectra of the light curve of PKS J1309+1154. These make the presence of the harmonic clear. The four panels are as follows: (a) The fundamental, with period $\sim 17.9$~yr,  is clearly visible over the full 46-year span of the observations.  However, the harmonic is not visible in this figure.  (b)  Here we have used the square root of the power, which enhances the harmonic relative to the fundamental. The harmonic is clearly visible up to 2012.5.  From 2012.5 on we have tested adding the OVRO data incrementally one year at a time, and again using the square root of the power.  (c) .  Here we see that the harmonic is clearly visible up to 2015. After 2015 the relative strength of the harmonic increases up to 2019. (d) This shows the harmonic strength increases with time compared to (c).  Hereafter the relative strength of the harmonic remains constant, but the width of the harmonic feature increases. These WWZ spectra demonstrate clearly that the harmonic feature is present and coherent over the whole 46-year light curve. This is consistent with the fact that the quadratic fits shown in Fig.~\ref{plt:J1309regions} in regions  10 and 11 continue to oscillate in phase with the harmonic.  We interpret this as showing that the underlying harmonic is indeed present, but  is somewhat masked by the variations that are not related to the sinusoidal component. The critical test will be to see whether the light curve of PKS J1309+1154 is approximately flat from 2026 to 2034.

\section{Jetted SMBHB Interpretation}\label{sec:interpretation}
\begin{figure}
    \centering
    \includegraphics[width=\linewidth]{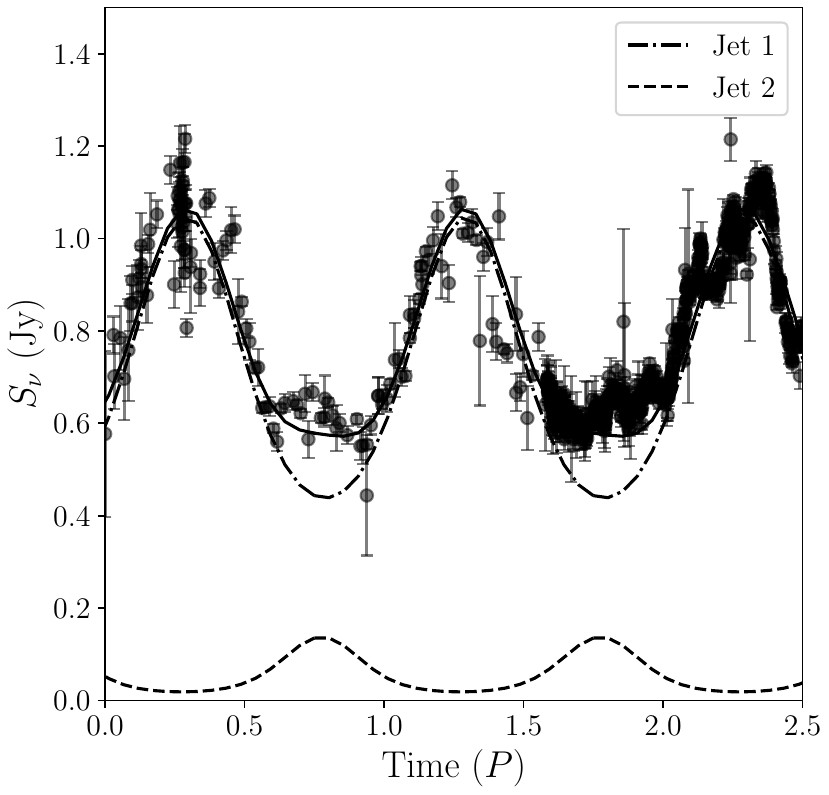}
    \caption{The proposed model for the de-trended PKS~J1309+1154 OVRO light curve using a modified version of the SMBHB jet model of Paper 4. Both black holes in the SMBHB have jets, but these are of unequal strength. The model jet parameters are listed in the text.}
    \label{fig:model}
\end{figure}
As in the cases of PKS 2131--021 and PKS J0805--0111 (Papers 1-3 and \citealt{2025arXiv250404278H}), one of the most natural explanations for a periodicity in PKS~J1309+1154 is an SMBHB with redshifted orbital period $P=17.9$ yr. The jets, which produce the beamed radio emission seen in these systems, can be dragged by the binary orbital motion, leading to variable Doppler beaming (\citealt{2017MNRAS.465..161S} and Paper 1).
Paper 4 proposed a model in which one black hole produces a jet, which is collimated by a sub-relativistic disk wind. The orbital motion causes the wind to form a helical channel through which the relativistic jet flows (see the bottom panel of Fig.~1 in Paper 4), and the redirection of the jet as it flows through the channel causes the change in beaming relative to the observer. The helix has wavelength $\lambda_h=\beta_wcP$, where $\beta_w$ is the speed of the wind-jet boundary and consequently the pattern speed of the helix. 

The jet is modeled as a synchrotron-emitting cone whose jet velocity direction is sinusoidally perturbed by an angle $\sim\beta_H/\beta_w$, where $\beta_H$ is the orbital speed (Paper 4). We assume a power-law synchrotron spectrum from radius $r_{\min}$ to $r_{\max}$, and calculate the radiative transfer of the radio emission through the jet to determine the flux density. The model parameters are the electromagnetic power $L_m$, the power in radiating particles $L_e$, the bulk Lorentz factor of the jet $\Gamma_j$, jet opening angle $\theta_j$, $\beta_w$, $\beta_H$, jet viewing inclination $i$, and the radiating electron/positron energy spectrum ranging from $\gamma_{e,\min}$ to $\gamma_{e,\max}$ with spectral index $p$. See Paper 4 for more details.

While this model can explain the multiwavelength observations of PKS J0805--0111, PKS J1309+1154, and PKS 2131--021, it is important to note that it does not produce a clear first harmonic, so the harmonic in PKS~J1309+1154 complicates the story. One possible explanation for the in-phase harmonic (or possible secondary maximum) is the emission from a weaker jet attached to the second black hole, which, in the model, amounts to adding the contribution of a second jet out of phase by $P/2$, assuming the orbit is circular. Additionally, to capture the effects of a harmonic, we cannot expand the variation in Doppler factor to lowest order in $\beta_H$. Instead, we must compute the full Doppler factor as a function of time. 

Following Paper 4, we write the redirected radial unit vector of a particular position within the open jet $\hat{r}$ as 
\begin{equation}
\label{eq:direction}
    \hat{r}(t) =\frac{1}{\sqrt{1+2\hat{r}_0\cdot\frac{\vec{\beta}_H}{\beta_w}+\frac{\beta_H^2}{\beta_w^2}}}\left(\hat{r}_0+ \frac{\vec{\beta}_H(t)}{\beta_w}\right),
\end{equation}
where $\hat{r}_0$ is the unperturbed radial unit vector, which corresponds to a fixed direction in the black hole frame. $\vec{\beta}_H$ introduces sinusoidal perturbation to $\hat{r}$. At a particular radius $r$ from the black hole, $\vec{\beta}_H$ in equation (\ref{eq:direction}) is delayed in phase from the instantaneous orbital velocity by $r/\lambda_h$. The local relativistic flow velocity in the observer frame is then $\vec{\beta}_j=\beta_j \hat{r}$, where $\beta_j$ is the velocity associated with $\Gamma_j$. The jet Doppler factor is 
\begin{equation}
\label{eq:Doppler}
    D(t)=\frac{1}{\Gamma_j\left(1-\beta_j\left(\hat{r}(t)\cdot\hat{n}\right)\right)},
\end{equation}
where $\hat{n}$ is the fixed observer direction. The flux density at observed frequency $\nu$ for each jet is
\begin{equation}
\begin{split}
\label{eq:fluxdensity_total}
S_\nu(t)&=(1+z_S)\,d_L^{-2}\iiint D(t)^{\frac{3+p}{2}}j^\prime_{\nu_z, \Omega} e^{-\tau_\nu(t)} dV,
\end{split}
\end{equation}
where $z_S$ is the source redshift, $d_L$ is the luminosity distance, $\tau_\nu(t)$ is the optical depth (which depends on $D(t)$), and $j^\prime_{\nu_z, \Omega}$ is the pitch-angle averaged emissivity in the comoving frame given by the right hand side of equation (2) in Paper 4. 
For volume element $dV=r^2 \sin\chi dr d\chi d\psi$, the integration bounds are $r\in[r_{\min}, r_{\max}]$, $\chi\in[0, \theta_j]$, and $\psi \in[0, 2\pi]$. The main difference between this treatment and that presented in Paper 4 is that here we treat the variation in $D$ and consequently $S_\nu$ non-perturbatively. We evaluate equation (\ref{eq:fluxdensity_total}) for two jets and sum them to get the total emission. 

In Fig.~\ref{fig:model}, we show a plausible model light curve for PKS~J1309+1154, including the contributions of each of the two jets. Since there is no reliable redshift for this source, we assume $z_S=1.3$ similar to PKS 2131--021 and PKS J0805--0111. We choose a nominal $i=4^\circ$ inclination. For the primary jet, we take $L_{e,1}=7\times10^{45} \text{ erg s}^{-1}\, (r/r_0)/(1+(r/r_0))$ (where $r_0=5$ ly),  $L_{m,1}=7\times10^{45} \text{ erg s}^{-1}\, 1/(1+(r/r_0))$, $\theta_{j,1}=0.05$, and $\Gamma_{j,1}=15$. The scaling is chosen so that $L_{j,1}=L_{m,1}+L_{e,1}$ remains constant. We assume the particle spectrum is initialized with $\gamma_{e,\min}=10$, $\gamma_{e,\max}=300$, and $p=1.2$ across the jet with normalization set by $L_{e,1}(r)$. The flatness in the spectrum makes it difficult to know the precise spectral index. These choices would be consistent with magnetic reconnection as the dominant electron/positron acceleration mechanism, although the acceleration mechanism remains unknown in these sources (Paper 4). We choose $\beta_{w,1}=0.9$, $\beta_{H,1}=0.02$, $r_{\min}$ so that $\tau_\nu(r_{\min})=20$, and $r_{\max}=r_{\min}+100 \,(\nu(1+z) /15 \text{ GHz})^{-1}$ ly. For the secondary jet, we keep all parameters the same except we take $L_{e,2}=1\times10^{45} \text{ erg s}^{-1}\, (r/r_0)/(1+(r/r_0))$, $L_{m,2}=1\times10^{45} \text{ erg s}^{-1}\, 1/(1+(r/r_0))$ and $\beta_{H,2}=0.4$ corresponding to mass ratio $q=0.5$. These parameters, particularly, $L_j$, $\Gamma_j$, $\theta_j$, and $r_{\max}$ differ from those used in PKS 2131--021 and PKS J0805--0111 (Paper 4). The changes made accommodate the sharper peaks seen in the 2.5 cycles of the primary jet.

The main parameter governing the relative height of the peaks is the relative values of $L_j$ between the two jets. Since we are only modeling the light curve at one observing frequency (15 GHz) and the wind speed cannot be well constrained. The quoted parameters are meant to be illustrative and can be more confidently constrained with multi-wavelength light curves and a more precise redshift.

As a consistency check, we can also estimate the corresponding mass ratio of the binary assuming for simplicity that the difference in luminosity was solely a result of different accretion rates onto the two black holes on orbital scales.
Under those assumptions using the a fit of mass accretion rate ratios from hydrodynamical simulations \citep{2023ARA&A..61..517L}, we then estimate a binary mass ratio $q \simeq 0.5$.

There are various issues with explaining the possible secondary maximum with a dual jet. If the opening angle $\theta_j>\beta_H/\beta_w$ for both jets or if $\beta_{w,1}\neq\beta_{w,2}$, the jets will crash into one another and likely merge into one jet. If this is the case, it is unlikely that there would be two stable jets to produce clean light curve peaks. There are other models in which one jet can produce the variation due to the internal structure of the jet. 

If there is a secondary maximum in the light curve of a source like J1309, then it is possible to account for it in the context of a single jet, in principle at least, by invoking a more complicated model of the emissivity than we have used so far. If we suppose that the jet rest frame emissivity is uniform on spherical surfaces within the jet, then it is easy to see that the observed flux will exhibit one maximum and one minimum within an orbital period. The flux, though, need not be sinusoidal. The first complication is to allow the jet Lorentz factor to vary with polar angle $\chi$ within the jet. A variation $\Gamma_j\propto (r\sin \chi)^{-1}$ can comply with the causality constraint and accommodate the presence of a sheath. Allowing $L_j$ to vary with $\chi$ can also lead to an edge-brightened jet. The second complication arises from the variation of the emissivity with jet azimuth $\psi$. In the model, the jet is confined and directed by the surrounding wind. Due to the orbital motion, individual jet elements are flowing into the wind asymmetrically, which could lead to the leading side possessing a higher emissivity than the trailing side (i.e. higher $L_j$ on the side of the jet moving into the wind). A model including these features may be able to produce the observed secondary maximum. A more complicated formalism, capable of exhibiting these effects, will be presented elsewhere.

We caution that, in principle, other potential explanation for harmonics in these sources have been proposed besides SMBHB motion, including accretion disk modes \citep{2013MNRAS.434.3487A}, or (first) harmonics associated with disk tearing and precession \citep{2023MNRAS.518.1656M}. It is also possible that some global jet mode may also produce this emission.  All of these alternative explanations raise potential questions about long-term period and harmonic stability. Of course, with only three cycles of the fundamental continued monitoring will be needed to determine the stability.

\section{Discussion}\label{sec:discussionl}

It is important to recognize that the emission at radio wavelengths is well known to originate in multiple regions along the jet (see, e.g., Papers 1 and 4). Therefore, not all of the emission in PKS J1309+1154 comes from the zone that is responsible for the sinusoidal variations. In particular (i) the regions giving rise to the long term trend are not those producing the sinusoidal variations, and (ii) the short term deviations from the sinusoidal emission originate in emission regions that are not related to the sinusoidal emission, as discussed in detail in Paper 1.

The combined UMRAO+OVRO light curve of PKS J1309+1154 and the detailed analysis in \S \ref{sec:J1309} amount, in our view, to a compelling argument that the periodicities related to the fundamental and harmonic are caused by an underlying periodicity in the blazar itself, and most likely in its central engine, and are not a random product of the steep slope of the PSD. We regard this evidence as conclusive for the following two reasons:
\begin{enumerate}
\item The least squares quadratic fits to the data in the eleven regions separated by the zero points in the first harmonic clearly lie on the same side of the fundamental curve as does the fundamental + harmonic curve. In other words, the data switches sides coherently with the harmonic in all eleven regions.  On the null hypothesis that the harmonic is not real, the data should be randomly distributed on either side of the fundamental curve, so the probability of this happening by chance has p-value = $4.9\times 10^{-4}$ --- i.e.  this result is significant at the  3.3$\sigma$ level.
\item The light curve displays two comparatively flat 8-year-long sections, from 1990 to 1998 and again from 2008 to 2016. The fundamental + harmonic curve likewise displays flat regions over these two 8-year windows, whereas the fundamental curve alone does not.
\end{enumerate}

For these reasons, we find the evidence for the harmonic compelling. This, of course, makes the evidence for the fundamental equally strong. This is the first time, to our knowledge, that such clear evidence for a fundamental periodicity and a harmonic periodicity has been found in a blazar light curve.

These results on PKS J1309+1154 provide conclusive evidence that some blazar light curves exhibit real, physical sinusoidal periodicities which are very likely produced in the central engine.  We regard the light curve of PKS J1309+1154 as providing the strongest evidence yet that some of the sinusoidal variations appearing in blazar light curves are real, and not a result of the steep slope of the PSD.  The fact that we now know that some sinusoidal variations are real greatly strengthens the case for the same interpretation of the light curves of PKS J0805--0111 \citep{2025ApJ...987..191D} and, in particular, of PKS 2131--021 (Papers 1 and 2), given that these both easily passed our $3\sigma$ test based on simulations. Furthermore, the fact that in PKS 2131--021 coherent sinusoidal variations are seen at optical wavelengths shows that this sinusoidal emission persists all the way back, from the radio-emitting regions, to a region very close to the central engine itself, where the emission must be very nearly in phase with the physical mechanism  in the central engine that is producing the periodicity, thus revealing the phase, as well as the period, of this mechanism.

The discovery of a compelling first harmonic in a blazar light curve opens up a new dimension to searches for SMBHB candidates, because it provides, for the first time,  a strong \textit{predictive\/} test of any periodicities that might be found.  While not all SMBHB candidates will necessarily have first harmonics, the fact that one has been found among the first three sinusoidally varying strong SMBHB candidates (PKS J0805-0111, PKS J1309+1154, and PKS 2131-021) to be found in radio light curves suggests that a significant fraction of SMBHB candidates found in radio blazar light curves are likely to have detectable first harmonics.

We estimate (Paper 2) that $\sim 1$ in 100 blazars is an SMBHB candidate. Thus, given that the South Pole Telescope (95 GHz, 150 GHz, 220 GHz) \citep{carlstrom/etal:2011,benson/etal:2014}, the Simons Observatory (27 GHz - 285 GHz) (SO; \citealt{so_collaboration:2019,so_collaboration:2025}), and the Vera Rubin Observatory (320-1060 nm) \citep{2023A&A...675A.163C} will be monitoring tens of thousands of blazars over at least a ten-year period, which is likely to be extended, we estimate that, applying the filter of a first harmonic search to identify SMBHB candidates in which the periodicity is proven,  they will identify scores, if not hundreds, of such candidates for which the redshift, the period, and the phase of the periodicities will be known. Thus will enable powerful searches for gravitational waves from these objects.  The remaining unknowns are the orbital velocities of the SMBHBs.  This will motivate intensive searches for optical and infrared spectral line variations due to orbital motion, which, if identified, would yield the masses of the two binary components. 

This information would then make possible a fully coherent search for gravitational waves from a network  of SMBHBs for which the periods, phases, and amplitudes of the expected gravitational waves from each individual SMBHB candidate would be known.  Since the directions, amplitudes, and phases, of all these individual gravitational wave signals would be known, they could all be summed \textit{in phase} with the correct amplitudes, making possible a fully coherent search for gravitational waves from the network of SMBHB candidates.

For the above reasons, we believe that the discovery of compelling evidence for a first harmonic in the combined UMRAO+OVRO light curve of the blazar PKS J1309+1154 is a major breakthrough in the quest for periodicities in blazars that are likely due to SMBHBs, making them promising targets for searches for gravitational waves using pulsar timing arrays.

\begin{acknowledgments}
This work is supported by NSF grants AST2407603 and AST2407604. We thank the California Institute of Technology and the Max Planck Institute for Radio Astronomy for supporting the  OVRO 40\,m program under extremely difficult circumstances over ten years (2014--2024) in the absence of agency funding for operation of the telescope. Without this private support this program would have ended in 2016.  We also thank all the volunteers who have enabled this work to be carried out.
Prior to~2016, the OVRO program was supported by NASA grants \hbox{NNG06GG1G}, \hbox{NNX08AW31G}, \hbox{NNX11A043G}, and \hbox{NNX13AQ89G} from~2006 to~2016 and NSF grants AST-0808050 and AST-1109911 from~2008 to~2014. The UMRAO program received support from NSF grants AST-8021250, AST-8301234, AST-8501093, AST-8815678, AST-9120224, AST-9421979, AST-9900723, AST-0307629, AST-0607523, and earlier NSF awards, and from NASA grants NNX09AU16G, NNX10AP16G, NNX11AO13G, and NNX13AP18G. Additional funding for the operation of UMRAO was provided by the University of Michigan. 
W.{}M.\ acknowledges support from ANID project Basal FB210003. A.S.  and R.B acknowledge support by a grant from the Simons Foundation (00001470,RB,AS). 
Y.{}D. and F.{}A.{}H.  acknowledge support through NASA under contract No. NNG08FD60C.
R.{}R.\ and B.{}M.\ and P.{}V.{}d.{}l.{}P.\ acknowledge support from ANID Basal AFB-170002, Núcleo Milenio TITANs (NCN2023\_002), CATA BASAL FB210003 and UdeC-VRID 2025001479INV.
T.{}H.\ acknowledges support from the Academy of Finland projects 317383, 320085, 345899, and 362571 and from the European Union ERC-2024-COG - PARTICLES - 101169986.
K.{}T.\ acknowledges the support by the TITAN ERA
Chair project (contract no. 101086741) within the Horizon Europe Framework
Program of the European Commission.
This research is partially funded by the European Union. Views and opinions expressed are, however, those of the author(s) only and do not necessarily reflect those of the European Union or the European Research Council Executive Agency. Neither the European Union nor the granting authority can be held responsible for them. V. P. is supported by an ERC grant, mw-atlas project no. 101166905. I.L is funded by the European Union ERC-2022-STG - BOOTES - 101076343.

\end{acknowledgments}

\facilities{OVRO:40m, UMRAO, NRAO:VLBA, ALMA, ZTF, WISE}

  \clearpage
    \appendix
\twocolumngrid
\section{Sine Wave Fitting}\label{sec:sine}

We fitted a model that includes two sine waves with periods $P$ and $P/2$ and a linear trend to the light curve of PKS~J1309+1154. In this model, the flux density is given by
\begin{align}
\begin{split}
S(t) &= S_0 + A_1 \sin(\phi (t) - \phi_{0,1}) \\
&+ A_2 \sin(2\phi(t) - \phi_{0,2}) + \beta (t - t_0) / 365.25,
\end{split}
\end{align}
where $\phi(t) = 2\pi(t-t_0)/P$ and $t_0 = 52,000$. Here, $P$ is the fundamental period, $A_1$ and $A_2$ are amplitudes of the fundamental and the harmonic, $\phi_{0,1}$ and $\phi_{0,2}$ are their phases at $t=t_0$, and $S_0$ is the mean flux density. We found the best-fit parameters by maximizing the following likelihood function:
\begin{equation}
\ln\mathcal{L} = -\frac{1}{2}\sum_i\left[\frac{(S_i - S(t_i))^2}{\sigma_i^2+\sigma_0^2} + \ln\left(\sigma_i^2+\sigma_0^2\right)\right],
\label{eq:likelihood}
\end{equation}
where $\sigma_0$ is a parameter that accounts for additional scatter in the data that is not captured by the original error bars. We used the Markov Chain Monte Carlo (MCMC) sampler by \citet{foreman2013} to derive the posterior distributions for all model parameters. The resulting best-fit values are reported in Table~\ref{tab:sine_parameters}.

\begin{table}[h]
\caption{Best-fit parameters of the sine-wave model for PKS~J1309+1154}
\label{tab:sine_parameters}
\begin{tabular}{lr}
\hline \hline
Parameter & Value\\
\hline
$P$ (days)       & $6550 \pm 19$          \\
$A_1$ (Jy)         & $0.2167 \pm 0.0022$    \\
$A_2$ (Jy)         & $0.0543 \pm 0.0022$    \\
$\phi_{0,1}$ (rad) & $-1.133 \pm 0.017$     \\
$\phi_{0,2}$ (rad) & $-0.372 \pm 0.055$     \\
$\beta$ (Jy/yr)    & $-0.00769 \pm 0.00018$ \\
$S_0$ (Jy)         & $0.7856 \pm 0.0027$    \\
$\sigma_0$ (Jy)    & $0.0440 \pm 0.0012$    \\
\hline
%$\ln\mathcal{L}_{\rm max}$ & 2570.4 & 2586.4 & 2449.3\\
%BIC & $-5085.4$ & $-5110.5$ & $-4850.1$\\
%\hline
\end{tabular}
\end{table}

\section{Quadratic fits}

We then fitted a quadratic polynomial to each region of the light curve, as defined in Table~\ref{tab:cycles}. We anchored the polynomial at the crossing points of the fundamental and the harmonic. Assuming that the polynomial crosses points $(x_1, S_1)$ and $(x_2, S_2)$, the flux density can be expressed as follows:
\begin{equation}
S(x) = S_1 + \frac{S_2-S_1}{x_2-x_1}(x-x_1) + p (x-x_1)(x-x_2),    
\end{equation}
where $p$ is a parameter and $x \equiv (t - t_1)/365.25$. The best-fit value of $p$ and its uncertainty were calculated by maximizing the likelihood function defined in equation.~(\ref{eq:likelihood}). The results are reported in Table~\ref{tab:quadratic}.

\begin{table*}
\caption{Best-fit parameters of the quadratic polynomial fits for PKS~J1309+1154}
\label{tab:quadratic}\begin{tabular}{crrrrrr}
\hline \hline
\multicolumn{1}{c}{Region} & \multicolumn{1}{c}{$t_1$} & \multicolumn{1}{c}{$t_2$} & \multicolumn{1}{c}{$S_1$} & \multicolumn{1}{c}{$S_2$} & \multicolumn{1}{c}{$p$} & \multicolumn{1}{c}{$\sigma_0$} \\
\hline
1 & 43619 & 45256 & 0.8355 &  1.1035 & $0.0000 \pm 0.0040$ & $0.0223 \pm 0.0159$\\
2 & 45256 & 46894 & 1.1035 &  1.0196 & $-0.0222 \pm 0.0032$ & $0.0812 \pm 0.0119$\\
3 & 46894 & 48531 & 1.0196 &  0.6828 & $0.0103 \pm 0.0050$ & $0.0735 \pm 0.0183$\\
4 & 48531 & 50169 & 0.6828 &  0.6976 & $0.0039 \pm 0.0033$ & $0.0422 \pm 0.0134$\\
5 & 50169 & 51806 & 0.6976 &  0.9655 & $0.0161 \pm 0.0031$ & $0.0494 \pm 0.0097$\\
6 & 51806 & 53444 & 0.9655 &  0.8817 & $-0.0156 \pm 0.0056$ & $0.0749 \pm 0.0229$\\
7 & 53444 & 55081 & 0.8817 &  0.5448 & $0.0123 \pm 0.0006$ & $0.0191 \pm 0.0017$\\
8 & 55081 & 56719 & 0.5448 &  0.5597 & $0.0009 \pm 0.0005$ & $0.0197 \pm 0.0014$\\
9 & 56719 & 58356 & 0.5597 &  0.8276 & $0.0047 \pm 0.0010$ & $0.0466 \pm 0.0028$\\
10 & 58356 & 59994 & 0.8276 &  0.7438 & $-0.0248 \pm 0.0011$ & $0.0593 \pm 0.0032$\\
11 & 59994 & 61631 & 0.7438 &  0.4069 & $0.0152 \pm 0.0024$ & $0.0550 \pm 0.0064$\\
\hline
\end{tabular}
\end{table*}

\begin{table*}
\caption{Best-fit parameters of the quadratic polynomial fits for PKS~J1309+1154}
\label{tab:quadratic2}\begin{tabular}{crrrrrr}
\hline \hline
\multicolumn{1}{c}{Region} & \multicolumn{1}{c}{$t_1$} & \multicolumn{1}{c}{$t_2$} & \multicolumn{1}{c}{$S_1$} & \multicolumn{1}{c}{$S_2$} & \multicolumn{1}{c}{$p$} & \multicolumn{1}{c}{$\sigma_0$} \\
\hline
1 & 45256 & 46894 & 0.9641 &  0.9141 & $-0.0229 \pm 0.0011$ & $0.0633 \pm 0.0029$\\
2 & 46894 & 48531 & 0.9141 &  0.6111 & $0.0138 \pm 0.0007$ & $0.0330 \pm 0.0018$\\
3 & 48531 & 50169 & 0.6111 &  0.6598 & $0.0017 \pm 0.0005$ & $0.0211 \pm 0.0013$\\
4 & 50169 & 51806 & 0.6598 &  0.9615 & $0.0061 \pm 0.0010$ & $0.0482 \pm 0.0026$\\
\hline
\end{tabular}
\end{table*}

%------------------------------------------------------------------------------
\bibliography{references}{}
\bibliographystyle{aasjournalv7}

%------------------------------------------------------------------------------

%------------------------------------------------------------------------------

%------------------------------------------------------------------------------

%------------------------------------------------------------------------------

\end{document}